\documentclass{article}
\usepackage{amsmath}

\input{tcilatex}
\begin{document}

\title{Cloning of Qubits of a Quantum Computer}
\author{V.N.Dumachev\thanks{%
E-mail: dum@comch.ru} \ S.V.Orlov \\
%EndAName
Voronezh Militia Institute, Ministry of Internal Affairs of the Russia}
\maketitle

\begin{abstract}
A system of unitary transformations providing two optimal copies of an
arbitrary input cubit is obtained. An algorithm based on classical Boolean
algebra and allowing one to find any unitary transformation realized by the
quantum CNOT operators is proposed.
\end{abstract}

It is known that an arbitrary quantum state

\begin{equation}
\left| {\psi }\right\rangle _{0}=\alpha \left| {0}\right\rangle _{0}+\beta
\left| {1}\right\rangle _{0},  \label{eq1}
\end{equation}

\noindent cannot be copied exactly (cloned). The no-cloning theorem was
proved in [1]. However, Buzek and Hillery [2] found a unitary transformation
entangling two qubits $\left| {\psi }\right\rangle _{12}=\left| {00}%
\right\rangle _{12}$\ with the input qubit $\left| {\psi }\right\rangle _{0} 
$ so that the output state has the form

\begin{equation}
\left| {\Psi ^{out}}\right\rangle =\left| {\Phi _{0}}\right\rangle
_{01}\left| {0}\right\rangle _{2}+\left| {\Phi _{1}}\right\rangle
_{01}\left| {1}\right\rangle _{2},  \label{eq2}
\end{equation}%
where

\begin{equation}  \label{eq3}
\begin{array}{l}
\left| {\Phi _{0}} \right\rangle = \sqrt {\frac{{1}}{{6}}} \left( {2\alpha
\left| {00} \right\rangle + \beta \left| {01} \right\rangle + \beta \left| {%
10} \right\rangle} \right), \\ 
\left| {\Phi _{1}} \right\rangle = \sqrt {\frac{{1}}{{6}}} \left( {2\beta
\left| {11} \right\rangle + \alpha \left| {01} \right\rangle + \alpha \left| 
{10} \right\rangle} \right).%
\end{array}%
\end{equation}

The reduced qubit density operators $\rho _{0}^{out}$, $\rho _{1}^{out}$ \
and $\rho _{2}^{out}$ \ at the output are related to the input density
operator $\rho ^{in}$ in as

\begin{equation*}
\rho _{0,1}^{out} = \frac{{5}}{{6}}\rho ^{in} + \frac{{1}}{{6}}\rho _{ \bot
}^{in} ,
\end{equation*}

\begin{equation*}
\rho _{2}^{out} = \frac{{2}}{{3}}\rho ^{in} + \frac{{1}}{{3}}\rho _{ \bot
}^{in} .
\end{equation*}

Here $\rho _{\bot }^{in}=\left| {\psi _{\bot }}\right\rangle
_{0}\left\langle {\psi _{\bot }}\right| $, where $\left| {\psi _{\bot }}%
\right\rangle _{0}=\alpha \left| {1}\right\rangle _{0}-\beta \left| {0}%
\right\rangle _{0}$ is the state orthogonal to the input state, $\alpha
=e^{i\varphi }sin\left( {\vartheta /2}\right) $, and $\beta =cos\left( {%
\vartheta /2}\right) $. The quality of obtained copies is specified by the
cloning accuracy F, which is determined by the overlap of the input and
output states [3]

\begin{equation*}
F = \frac{{1}}{{4\pi} }\int\limits_{0}^{2\pi} {d\varphi} \int\limits_{0}^{%
\pi} {\left\langle {\psi ^{in}} \right|\rho ^{out}\left| {\psi ^{in}}
\right\rangle sin\vartheta d\vartheta} .
\end{equation*}

Thus, the output qubits $\rho _{0}^{out}$ and $\rho _{1}^{out}$ consist of $%
\frac{{5}}{{6}}$ fraction of the input qubit $\rho ^{in}$ in and $\frac{{1}}{%
{6}}$ fraction of an admixture. The qubit $\left| {\psi }\right\rangle _{2}$
is auxiliary and called cloning. Gisin and Massar [4] proved analytically
that representation (3) of the output qubits is optimal, i.e., maximizes the
average accuracy of the correspondence between the input and output qubits.

The sequence of actions for cloning qubits is represented in the form of a
universal quantum cloning machine. For its operation, it is necessary to
prepare preliminary the entangled state of two qubits

\begin{equation}  \label{eq4}
\left| {\Psi ^{prep}} \right\rangle = C_{1} \left| {00} \right\rangle +
C_{2} \left| {01} \right\rangle + C_{3} \left| {10} \right\rangle + C_{4}
\left| {11} \right\rangle
\end{equation}

\noindent by applying unitary operators to zeroth qubits:

\begin{equation*}
\left| {\Psi ^{prep}}\right\rangle =R_{1}\left( {\theta _{3}}\right)
P_{21}R_{2}\left( {\theta _{2}}\right) P_{12}R_{1}\left( {\theta _{1}}%
\right) \left| {00}\right\rangle _{12}.
\end{equation*}%
Here,

\begin{equation*}
R\left( {\theta }\right) =\left( 
\begin{array}{cc}
\cos \theta  & -\sin \theta  \\ 
\sin \theta  & \cos \theta 
\end{array}%
\right) 
\end{equation*}%
is the turning operator of a qubit and

\begin{equation}
P_{12}\left| {x,y}\right\rangle =\left| {x,x\oplus y}\right\rangle
\label{eq5}
\end{equation}%
is the CNOT operator, where $\oplus $ is the modulus-2 summation. The
resulting set of equations

\begin{equation}  \label{eq6}
\begin{array}{l}
cos\theta _{1} cos\theta _{2} cos\theta _{3} + sin\theta _{1} sin\theta _{2}
sin\theta _{3} = C_{1} \\ 
sin\theta _{1} cos\theta _{2} cos\theta _{3} - cos\theta _{1} sin\theta _{2}
sin\theta _{3} = C_{2} \\ 
cos\theta _{1} cos\theta _{2} sin\theta _{3} - sin\theta _{1} sin\theta _{2}
cos\theta _{3} = C_{3} \\ 
cos\theta _{1} sin\theta _{2} cos\theta _{3} + sin\theta _{1} cos\theta _{2}
sin\theta _{3} = C_{4}%
\end{array}%
\end{equation}

\noindent has the solution

\begin{equation*}
cos^{2}\theta _{1} = \frac{{C_{2}^{2} - C_{3}^{2}} }{{1 - 2C_{3}^{2} -
2C_{4}^{2}} } + cos^{2}\theta _{3} \frac{{1 - 2C_{2}^{2} - 2C_{4}^{2}} }{{1
- 2C_{3}^{2} - 2C_{4}^{2}} },
\end{equation*}

\begin{equation}  \label{eq7}
cos^{2}\theta _{2} = \frac{{C_{3}^{2} + C_{4}^{2} - cos^{2}\theta _{3}} }{{1
- 2cos^{2}\theta _{3}} },
\end{equation}

\begin{equation*}
cos^{2}\theta _{3}=\frac{{1}}{{2}}\left( {1\pm \frac{{1-2C_{3}^{2}-2C_{4}^{2}%
}}{{1-4\left( {C_{1}^{2}C_{4}^{2}+C_{2}^{2}C_{3}^{2}}\right) }}\sqrt{%
1-4\left( {C_{1}^{2}C_{4}^{2}+C_{2}^{2}C_{3}^{2}}\right)
+8C_{1}C_{2}C_{3}C_{4}}}\right) .
\end{equation*}

\bigskip \textbf{Table 1.}

\bigskip 
\begin{tabular}{|c|c|c|c|c|c|c|}
\hline
N & $\left( C_{1},C_{2},C_{3},C_{4}\right) $ & cos$^{2}\theta _{1}$ & cos$%
^{2}\theta _{2}$ & cos$^{2}\theta _{3}$ & sign$\left( \cos \theta _{i},\sin
\theta _{i}\right) $ & $\left| {\Psi ^{out}}\right\rangle $ \\ \hline
1 & $\dfrac{1}{\sqrt{6}}\left( 2,1,1,0\right) $ & $\dfrac{1}{2}\left( 1\mp 
\dfrac{1}{\sqrt{2}}\right) $ & $1\mp \dfrac{\sqrt{2}}{3}$ & $\frac{1}{2}%
\left( 1\mp \dfrac{1}{\sqrt{2}}\right) $ & $\left. 
\begin{array}{c}
\left( ---,+++\right) \\ 
\left( +++,+-+\right)%
\end{array}%
\right. $ & $\left. 
\begin{array}{c}
P_{21}P_{02}P_{10}\left| {\Psi ^{in}}\right\rangle \\ 
P_{12}P_{20}P_{01}\left| {\Psi ^{in}}\right\rangle%
\end{array}%
\right. $ \\ \hline
2 & $\dfrac{1}{\sqrt{6}}\left( 2,1,0,1\right) $ & $\dfrac{1}{2}\left( 1\mp 
\dfrac{1}{\sqrt{5}}\right) $ & $\dfrac{1}{2}\left( 1\mp \dfrac{\sqrt{5}}{3}%
\right) $ & $\dfrac{1}{2}\left( 1\mp \dfrac{2}{\sqrt{5}}\right) $ & $\left. 
\begin{array}{c}
\left( +++,-+-\right) \\ 
\left( +++,+++\right)%
\end{array}%
\right. $ & $\left. 
\begin{array}{c}
P_{21}P_{10}P_{02}\left| {\Psi ^{in}}\right\rangle \\ 
P_{10}P_{20}P_{02}P_{01}\left| {\Psi ^{in}}\right\rangle%
\end{array}%
\right. $ \\ \hline
3 & $\dfrac{1}{\sqrt{6}}\left( 2,0,1,1\right) $ & $\dfrac{1}{2}\left( 1\mp 
\dfrac{1}{\sqrt{5}}\right) $ & $\dfrac{1}{2}\left( 1\mp \dfrac{\sqrt{5}}{3}%
\right) $ & $\dfrac{1}{2}\left( 1\mp \dfrac{1}{\sqrt{5}}\right) $ & $\left. 
\begin{array}{c}
\left( +++,-+-\right) \\ 
\left( +++,+++\right)%
\end{array}%
\right. $ & $\left. 
\begin{array}{c}
P_{12}P_{01}P_{20}\left| {\Psi ^{in}}\right\rangle \\ 
P_{01}P_{02}P_{20}P_{10}\left| {\Psi ^{in}}\right\rangle%
\end{array}%
\right. $ \\ \hline
4 & $\dfrac{1}{\sqrt{6}}\left( 1,2,1,0\right) $ & $\dfrac{1}{2}\left( 1\pm 
\dfrac{1}{\sqrt{5}}\right) $ & $\dfrac{1}{2}\left( 1\mp \dfrac{\sqrt{5}}{3}%
\right) $ & $\dfrac{1}{2}\left( 1\mp \dfrac{2}{\sqrt{5}}\right) $ & $\left. 
\begin{array}{c}
\left( ---,+++\right) \\ 
\left( +++,+-+\right)%
\end{array}%
\right. $ & $\left. 
\begin{array}{c}
P_{20}P_{10}P_{01}P_{0\overline{2}}\left| {\Psi ^{in}}\right\rangle \\ 
P_{21}P_{10}P_{0\overline{2}}\left| {\Psi ^{in}}\right\rangle%
\end{array}%
\right. $ \\ \hline
5 & $\dfrac{1}{\sqrt{6}}\left( 1,2,0,1\right) $ & $\dfrac{1}{2}\left( 1\pm 
\dfrac{1}{\sqrt{2}}\right) $ & $1\mp \dfrac{\sqrt{2}}{3}$ & $\dfrac{1}{2}%
\left( 1\mp \dfrac{1}{\sqrt{2}}\right) $ & $\left. 
\begin{array}{c}
\left( +++,-+-\right) \\ 
\left( +++,+++\right)%
\end{array}%
\right. $ & $\left. 
\begin{array}{c}
P_{12}P_{\overline{2}0}P_{01}\left| {\Psi ^{in}}\right\rangle \\ 
P_{21}P_{0\overline{2}}P_{10}\left| {\Psi ^{in}}\right\rangle%
\end{array}%
\right. $ \\ \hline
6 & $\dfrac{1}{\sqrt{6}}\left( 1,1,2,0\right) $ & $\dfrac{1}{2}\left( 1\mp 
\dfrac{2}{\sqrt{5}}\right) $ & $\dfrac{1}{2}\left( 1\mp \dfrac{\sqrt{5}}{3}%
\right) $ & $\dfrac{1}{2}\left( 1\pm \dfrac{1}{\sqrt{5}}\right) $ & $\left. 
\begin{array}{c}
\left( ---,+++\right) \\ 
\left( +++,+-+\right)%
\end{array}%
\right. $ & $\left. 
\begin{array}{c}
P_{01}P_{02}P_{20}P_{\overline{1}0}\left| {\Psi ^{in}}\right\rangle \\ 
P_{12}P_{0\overline{1}}P_{20}\left| {\Psi ^{in}}\right\rangle%
\end{array}%
\right. $ \\ \hline
7 & $\dfrac{1}{\sqrt{6}}\left( 1,1,0,2\right) $ & $\dfrac{1}{2}\left( 1\pm 
\dfrac{2}{\sqrt{5}}\right) $ & $\dfrac{1}{2}\left( 1\mp \dfrac{\sqrt{5}}{3}%
\right) $ & $\dfrac{1}{2}\left( 1\pm \dfrac{1}{\sqrt{5}}\right) $ & $\left. 
\begin{array}{c}
\left( +++,-+-\right) \\ 
\left( +++,+++\right)%
\end{array}%
\right. $ & $\left. 
\begin{array}{c}
P_{12}P_{0\overline{1}}P_{\overline{2}0}\left| {\Psi ^{in}}\right\rangle \\ 
P_{01}P_{02}P_{\overline{2}0}P_{\overline{1}0}\left| {\Psi ^{in}}%
\right\rangle%
\end{array}%
\right. $ \\ \hline
8 & $\dfrac{1}{\sqrt{6}}\left( 1,0,2,1\right) $ & $\dfrac{1}{2}\left( 1\mp 
\dfrac{1}{\sqrt{2}}\right) $ & $1\mp \dfrac{\sqrt{2}}{3}$ & $\dfrac{1}{2}%
\left( 1\pm \dfrac{1}{\sqrt{2}}\right) $ & $\left. 
\begin{array}{c}
\left( +++,-+-\right) \\ 
\left( +++,+++\right)%
\end{array}%
\right. $ & $\left. 
\begin{array}{c}
P_{21}P_{02}P_{\overline{1}0}\left| {\Psi ^{in}}\right\rangle \\ 
P_{12}P_{20}P_{0\overline{1}}\left| {\Psi ^{in}}\right\rangle%
\end{array}%
\right. $ \\ \hline
9 & $\dfrac{1}{\sqrt{6}}\left( 1,0,1,2\right) $ & $\dfrac{1}{2}\left( 1\pm 
\dfrac{1}{\sqrt{5}}\right) $ & $\dfrac{1}{2}\left( 1\mp \dfrac{\sqrt{5}}{3}%
\right) $ & $\dfrac{1}{2}\left( 1\pm \dfrac{2}{\sqrt{5}}\right) $ & $\left. 
\begin{array}{c}
\left( +++,-+-\right) \\ 
\left( +++,+++\right)%
\end{array}%
\right. $ & $\left. 
\begin{array}{c}
P_{21}P_{\overline{1}0}P_{0\overline{2}}\left| {\Psi ^{in}}\right\rangle \\ 
P_{20}P_{10}P_{0\overline{1}}P_{0\overline{2}}\left| {\Psi ^{in}}%
\right\rangle%
\end{array}%
\right. $ \\ \hline
10 & $\dfrac{1}{\sqrt{6}}\left( 0,1,1,2\right) $ & $\dfrac{1}{2}\left( 1\pm 
\dfrac{1}{\sqrt{2}}\right) $ & $1\mp \dfrac{\sqrt{2}}{3}$ & $\dfrac{1}{2}%
\left( 1\pm \dfrac{1}{\sqrt{2}}\right) $ & $\left. 
\begin{array}{c}
\left( ---,+++\right) \\ 
\left( +++,+-+\right)%
\end{array}%
\right. $ & $\left. 
\begin{array}{c}
P_{21}P_{0\overline{2}}P_{\overline{1}0}\left| {\Psi ^{in}}\right\rangle \\ 
P_{12}P_{\overline{2}0}P_{0\overline{1}}\left| {\Psi ^{in}}\right\rangle%
\end{array}%
\right. $ \\ \hline
11 & $\dfrac{1}{\sqrt{6}}\left( 0,1,2,1\right) $ & $\dfrac{1}{2}\left( 1\mp 
\dfrac{1}{\sqrt{5}}\right) $ & $\dfrac{1}{2}\left( 1\mp \dfrac{\sqrt{5}}{3}%
\right) $ & $\dfrac{1}{2}\left( 1\pm \dfrac{2}{\sqrt{5}}\right) $ & $\left. 
\begin{array}{c}
\left( ---,+++\right) \\ 
\left( +++,+-+\right)%
\end{array}%
\right. $ & $\left. 
\begin{array}{c}
P_{21}P_{\overline{1}0}P_{02}\left| {\Psi ^{in}}\right\rangle \\ 
P_{20}P_{10}P_{0\overline{1}}P_{02}\left| {\Psi ^{in}}\right\rangle%
\end{array}%
\right. $ \\ \hline
12 & $\dfrac{1}{\sqrt{6}}\left( 0,2,1,1\right) $ & $\dfrac{1}{2}\left( 1\pm 
\dfrac{2}{\sqrt{5}}\right) $ & $\dfrac{1}{2}\left( 1\mp \dfrac{\sqrt{5}}{3}%
\right) $ & $\dfrac{1}{2}\left( 1\mp \dfrac{1}{\sqrt{5}}\right) $ & $\left. 
\begin{array}{c}
\left( ---,+++\right) \\ 
\left( +++,+-+\right)%
\end{array}%
\right. $ & $\left. 
\begin{array}{c}
P_{12}P_{01}P_{\overline{2}0}\left| {\Psi ^{in}}\right\rangle \\ 
P_{01}P_{02}P_{\overline{2}0}P_{10}\left| {\Psi ^{in}}\right\rangle%
\end{array}%
\right. $ \\ \hline
\end{tabular}

\newpage

\bigskip \FRAME{ftbpFU}{353.1875pt}{208.125pt}{0pt}{\Qcb{Scheme of the
optimized variant of the Buzek--Hillery cloning machine. Only two CNOT
operators participating in the production of the output state $\left| {\Psi }%
\right\rangle _{012}^{out}$ from $\left| {\protect\psi }\right\rangle _{0}$
and $\left| {\Psi }\right\rangle ^{prep}$ \ involve the input qubit $\left| {%
\protect\psi }\right\rangle _{0}$ . The arrows point to the goal qubit of
the CNOT operator.}}{}{Figure}{\special{language "Scientific Word";type
"GRAPHIC";display "USEDEF";valid_file "T";width 353.1875pt;height
208.125pt;depth 0pt;original-width 529.25pt;original-height
353.0625pt;cropleft "0";croptop "1";cropright "1";cropbottom
"0";tempfilename 'dum05.ps';tempfile-properties "XPR";}}

At the second stage, the quantum cloning machine mixes input qubit (1) with
prepared state (4):

\begin{equation}
\left| {\Psi ^{in}}\right\rangle =\left| {\psi }\right\rangle _{0}\left| {%
\Psi ^{prep}}\right\rangle =\alpha \left( {C_{1}\left| {000}\right\rangle
+C_{2}\left| {001}\right\rangle +C_{3}\left| {010}\right\rangle +C_{4}\left| 
{011}\right\rangle }\right) +\beta \left( {C_{1}\left| {100}\right\rangle
+C_{2}\left| {101}\right\rangle +C_{3}\left| {110}\right\rangle +C_{4}\left| 
{111}\right\rangle }\right) ,  \label{eq8}
\end{equation}

\noindent so as to obtain optimal state (2) at the output

\begin{equation}
\left| {\Psi ^{out}}\right\rangle =\left| {\Phi _{0}}\right\rangle
_{01}\left| {0}\right\rangle _{2}+\left| {\Phi _{1}}\right\rangle
_{01}\left| {1}\right\rangle _{2}=\sqrt{\frac{{1}}{{6}}}\left( {2\alpha
\left| {000}\right\rangle +\beta \left| {010}\right\rangle +\beta \left| {100%
}\right\rangle +2\beta \left| {111}\right\rangle +\alpha \left| {011}%
\right\rangle +\alpha \left| {101}\right\rangle }\right)  \label{eq9}
\end{equation}

Comparing Eqs. (3) and (9), we obtain only 12 different combinations of
admissible parameters $C_{1},C_{2},C_{3},$ and $C_{4}$, for which solution
(7) gives the angles for operators $R_{1}\left( {\theta _{1}}\right) $, $%
R_{2}\left( {\theta _{2}}\right) $, $R_{1}\left( {\theta _{3}}\right) $ (see
columns 2--5 in Table 1). The sixth column of Table 1 shows the signs of the
rotation angles of $R\left( {\theta }\right) $.

Now, we obtain transformations converting input state (8) (with known $%
C_{1},C_{2},C_{3},$ and $C_{4}$) to output state (9) only in terms of CNOT
operators (5). We represent the total transformation operator as

\begin{equation*}
\left| {\Psi ^{out}} \right\rangle _{xyz} = P\left( {x,y,z} \right)\left| {%
\Psi ^{in}} \right\rangle _{xyz} = \left| {p_{1} \left( {x,y,z}
\right),p_{2} \left( {x,y,z} \right),p_{3} \left( {x,y,z} \right)}
\right\rangle
\end{equation*}

\noindent where $p_{i}\left( {x,y,z}\right) $ are the logical functions of
three Boolean variables.

Let us find this function for the first row of Table 1.

\bigskip \textbf{Table 1. }Truth table for functions $p_{i}$

\begin{center}
\begin{tabular}{|c|c|c|c|c|c|}
\hline
$x$ & $y$ & $z$ & $p_{1}$ & $p_{2}$ & $p_{3}$ \\ \hline
0 & 0 & 0 & 0 & 0 & 0 \\ \hline
0 & 0 & 1 & 0 & 1 & 1 \\ \hline
0 & 1 & 0 & 1 & 0 & 1 \\ \hline
0 & 1 & 1 & * & * & * \\ \hline
1 & 0 & 0 & 1 & 1 & 1 \\ \hline
1 & 0 & 1 & 1 & 0 & 0 \\ \hline
1 & 1 & 0 & 0 & 1 & 0 \\ \hline
1 & 1 & 1 & * & * & * \\ \hline
\end{tabular}

Here, asterisks mean arbitrary values.
\end{center}

\qquad Since the CNOT operator can realize only linear Boolean functions,
only two of eight different combinations are suitable. For one of them, we
represent the set of disjunctive normal forms in terms of the Zhegalkin
polynomials:

\begin{equation}
\begin{array}{l}
p_{1}=\bar{x}\&y\&\bar{z}\vee \bar{x}\&y\&z\vee x\&\bar{y}\&\bar{z}\vee x\&%
\bar{y}\&z=x\oplus y, \\ 
p_{2}=\bar{x}\&\bar{y}\&z\vee \bar{x}\&y\&z\vee x\&\bar{y}\&\bar{z}\vee
x\&y\&\bar{z}=x\oplus z, \\ 
p_{3}=\bar{x}\&\bar{y}\&z\vee \bar{x}\&y\&\bar{z}\vee x\&\bar{y}\&\bar{z}%
\vee x\&y\&z=x\oplus y\oplus z.%
\end{array}%
.  \label{eq10}
\end{equation}%
Then,

\begin{equation*}
\left| {\Psi ^{out}}\right\rangle _{xyz}=\left| {p_{1},p_{2},p_{3}}%
\right\rangle =\left| {x\oplus y,x\oplus z,x\oplus y\oplus z}\right\rangle
=P_{21}P_{02}P_{10}\left| {\Psi ^{in}}\right\rangle _{xyz}.
\end{equation*}%
The other rows in Table 1 are filled similarly. To describe the CNOT
operator with inversion, we introduce the notation

\begin{equation*}
P_{1\bar {2}} \left| {x,y} \right\rangle = P_{12} \left| {x,\bar {y}}
\right\rangle = \left| {x,x \oplus \bar {y}} \right\rangle = P_{12} R_{2}
\left( {\frac{{\pi} }{{2}}} \right)\left| {x,y} \right\rangle ,
\end{equation*}

where $R\left( {\frac{{\pi }}{{2}}}\right) =\left( 
\begin{array}{cc}
0 & 1 \\ 
-1 & 0%
\end{array}%
\right) $ is the NOT operation.

The lower half of row 2 in Table 1 describes the operation of the
Buzek--Hillery quantum cloning machine [5], whereas the upper half of row 2,
describes its optimized variant. It is seen that the output state $\left| {%
\Psi ^{out}}\right\rangle $ can be obtained by three CNOT transformations,
only two of which involve the input qubit (see figure). The equatorial
qubits of the first row were studied in [6] without discussing the method of
their production.

In summary, we obtained the set of unitary transformations producing two
copies of an arbitrary input qubit. This transformation is optimal, because
it maximizes the average accuracy of correspondence between the input and
output qubits. The algorithm allowing one to find any unitary transformation
realized by the quantum CNOT operators is proposed on the basis of classical
Boolean algebra.

\begin{center}
\textbf{REFERENCES}
\end{center}

1. W. K. Wooters and W. H. Zuker, Nature 299, 802 (1982).

2. V. Buzek and M. Hillery, Phys. Rev. A 54, 1844 (1996).

3. R. F. Werner, quant-ph/9804001.

4. N. Gisin and S. Massar, Phys. Rev. Lett. 79, 2153 (1997).

5. V. Buzek, S. L. Braunstein, M. Hillery, and D. Bru, Phys. Rev. A 56, 3446
(1997).

6. Heng Fan, Keiji Matsumoto, and Xiang-Bin Wang, quant-ph/0101101.

\bigskip

Translated by R. Tyapaev

\end{document}